# Comment on: "Observation of 23 Supernovae that exploded <300 pc from Earth during the past 300 kyr" by R.B. Firestone

Adrian L. Melott[1], Ilya G. Usoskin[2], Gennady A. Kovaltsov[3], and Claude M. Laird[4]

1. Department of Physics and Astronomy, University of Kansas, USA

2. ReSoLVE Center of Excellence and Sodankylä Geophysical Observatory (Oulu unit) University of Oulu, Finland

3. Ioffe Physical-Technical Institute, St.Petersburg, Russia

4. Retired, University of Kansas, USA

ABSTRACT: We show that Firestone (2014) contains numerous errors in the application of past research. F14 overestimates likely nitrate and $^{14}$C production from moderately nearby supernovae by about four orders of magnitude. Moreover, the results are based on wrongly selected (obsolete) nitrate and $^{14}$C datasets. The use of correct and up-to-date datasets does not confirm the claimed results. The claims in the paper are invalidated.

*Introduction: Ionizing Radiation from Supernovae*

Firestone (2014; hereafter F14) bases arguments for abundant moderately nearby supernovae (SNe) on data of measured cosmogenic isotope deposition and nitrate accumulation in terrestrial archives. The claimed rate of 23 supernovae within 300 pc of the Earth within the last 300 kyr would exceed the average galactic rate by a factor of 4, so the claim is suspicious if only on this basis. The average galactic rate has about 2 supernovae per Myr within 100 pc (Fields 2004); due to the geometry of the galactic disk one would expect about 20 per Myr within 300 pc, or only 6 within the last 300 kyr. Of course, the Earth may lie in an unusually active region of the Galaxy, but such claims bear further examination.

We examine these claims here. The F14 computations depend upon the ionizing radiation (viz. hard X-, gamma-rays and cosmic rays) fluence from the supernovae. F14 deduces, for example, that there is as much energy as $2 \times 10^{49}$ erg for the initial burst of ionizing radiation. This is more typical of the total electromagnetic radiation output (including visible light) of a supernova, and considerably higher than the modern measurement of X-rays (Soderberg et al. 2008), which lies around $2 \times 10^{46}$ erg, with no gamma-rays detected. Over a period of months, X-ray emissions continue at a lower flux, accumulating as much as $10^{47}$ erg (Melott and Thomas 2011), although rare extreme outliers may produce two orders of magnitude more (Levan et al. 2013). There is a kinetic energy component of order $10^{51}$ erg; this may be taken as an upper limit to the possible energy in cosmic rays. Of course the photon transport to Earth is at the speed of light, but the cosmic rays have diffusive transport, taking hundreds to thousands of years longer for the cases we will consider. The PeV cosmic rays would arrive in perhaps 300 yr for a 100 pc distant supernova. But most would arrive later (e.g., for a SN at 250 pc distance the maximum of cosmic rays would arrive with a 4-40 kyr delay, using the mean free path of cosmic rays in the interstellar medium as 2.5-25 pc, see, e.g., Lingenfelter & Ramaty, 1970), and arriving over a similar time, and their time profile would have a typical shape of a diffusive propagation wave. There is no observed evidence for such a wave.

*Nitrates*

Ionizing radiation breaks the triple bond of $N_2$, making possible the synthesis of oxides of nitrogen in the atmosphere, which are normally present at low abundance. Nitrate peaks in ice cores have been proposed as signatures of supernovae, and F14 considers this question. Historical supernovae are used as examples, including in F14. It is now possible, given detailed numerical simulations (e.g. Thomas et al. 2007; Melott and Thomas 2011) to compute the nitrate deposition from ionizing radiation onto the

Earth. For X-rays and gamma-rays, $10^{-4}$ ng/erg is a good estimate of the global average in the X-ray regime, with no strong dependence upon the time development of the radiation, beyond simple causality (Ejzak et al. 2007).

Nitrate deposition at these low fluxes will scale nearly linearly with ionizing fluence at the Earth. Let us examine the historical supernovae, and parameterize the expected nitrate deposition based on their X-ray fluence and distance, using simply the inverse square law. The expected nitrate deposition from supernova X-rays is of order

d = 1 ng/cm² $(R/100 \text{ pc})^{-2}$ $(F/10^{46} \text{ erg})$,

where F is the total fluence of ionizing radiation for a SN at distance R.

F14 quotes Dreschhoff and Laird (2006) regarding evidence from historical supernovae. The following table shows the measured and expected nitrate deposition in the GISP2-H ice core from Summit, Greenland, assuming $10^{46}$ erg fluence. Distances are from Green (2004). These peaks include 1-2 months of deposition. By including the months of extended X-ray emissions the expected numbers may be increased, but still are four orders of magnitude too small to account for nitrate peaks speculatively associated with the historical supernovae.

| DATE | EVENT | DISTANCE | Nitrate measured ng/cm² | Nitrate expected ng/cm² |
|---|---|---|---|---|
| 1573/74 | Tycho | 2.3 kpc | 177 | .0019 |
| 1605 | Kepler | 2.9 kpc | 266 | .0012 |
| 1667 | Cas A? | 3.4 kpc | 150 | .0009 |
| 1700 | Cas A? | 3.4 kpc | 218 | .0009 |

Other historical supernova nitrate spikes cited by F14 were from Rood et al. (1979). The 1974 South Pole ice core cited by F14 was the first core from this site analyzed for nitrate and the conclusions of Rood et al. (1979) have been generally discredited. This assertion is based on Dreschhoff et al. (1983), who retracted the results after a second South Pole ice core was drilled in 1978 and found that most of the nitrate spikes in question could be attributed to "artifacts of contamination." They concluded, "While we cannot reject totally the idea that supernovae may be detectable in the nitrate signal, it is clear that the extreme spikes did not result from this source." This second South Pole core, along with an ice core from Vostok Station were cited in Dreschhoff and Laird (2006), however, the nitrate fluences above background (roughly 600 ng/cm2 for South Pole and Vostok samples) hypothesized as due to SNe are too large by more than 5 orders of magnitude to match predictions.

Photons from the 19 additional supernovae "observed" by F14 at distances of 100-300 pc could be expected to produce similar amounts of nitrate deposition to those "expected" in our Table, far below the noise level in these measurements. If the X-ray fluence were closer to the $10^{49}$ erg suggested by F14, which exceeds most supernovae, they would still be far too small to account for the measured nitrate.

The cosmic ray flux will arrive over hundreds to thousands of years, and may take a substantial fraction of the kinetic energy of the supernova; using the recent consensus value for the efficiency of conversion of bulk kinetic energy to cosmic rays of order 10%, we adopt $10^{50}$ erg as a typical value for the injection of cosmic rays into the interstellar medium. The arrival will be energy-dependent (e.g. Erlykin & Wolfendale 2010) with the highest energy cosmic rays arriving first and an extended tail of lower energy ones. The aggregate energy incident upon the Earth in cosmic rays will be of order $10^8$ erg/cm$^2$ for an 100 pc event. This would give a small, very extended, excess nitrate deposition which would be challenging to measure.

### *Carbon-14*

The primary argument of F14 is based on $^{14}$C variation. However, several crucial errors have been made here.

First, F14 analyzed data shown in his Fig. 2 to claim a saw-tooth structure with several peaks and decays. That Figure is a composite of two datasets – INTCAL04 (Reimer et al., 2003) for the age range 0-26 kyr age, and Hughen et al.(2004) for the age older than 26 kyr, the latter being arbitrarily lifted up by 22.5 % to match INTCAL04 at 26 kyrs ago. One can see that the two pieces do not match each other in the most recent well dated part, implying that the 22.5% offset is wrong. Moreover, these datasets are outdated. Hughen et al. have later strongly revised their dataset (Hughen et al., 2006, see Fig. 5 there), so that Fig.2b of F14 is dramatically modified, no longer showing the saw-tooth structure. The use of INTCAL04 is also not valid. The INTCAL series has been greatly updated recently with INTCAL09 and INTCAL13 (Reimer et al., 2013) officially released. It is important that the dataset of Hughen et al. (2006) is explicitly included in INTCAL13. The time series of Δ$^{14}$C for INTCAL13 shown in Fig. 1 has little in common with the dataset used by F14. In particular, there are no spikes ca. 18 and 22 kyrs ago. There are also no saw-tooth structures with exponential "decays" before 26 kyr ago (Fig. 2b in F14, 2014). The variability beyond 26 kyr is much smaller than claimed by F14 and can be explained by the climate and geomagnetic field variability. This invalidates F14's claims.

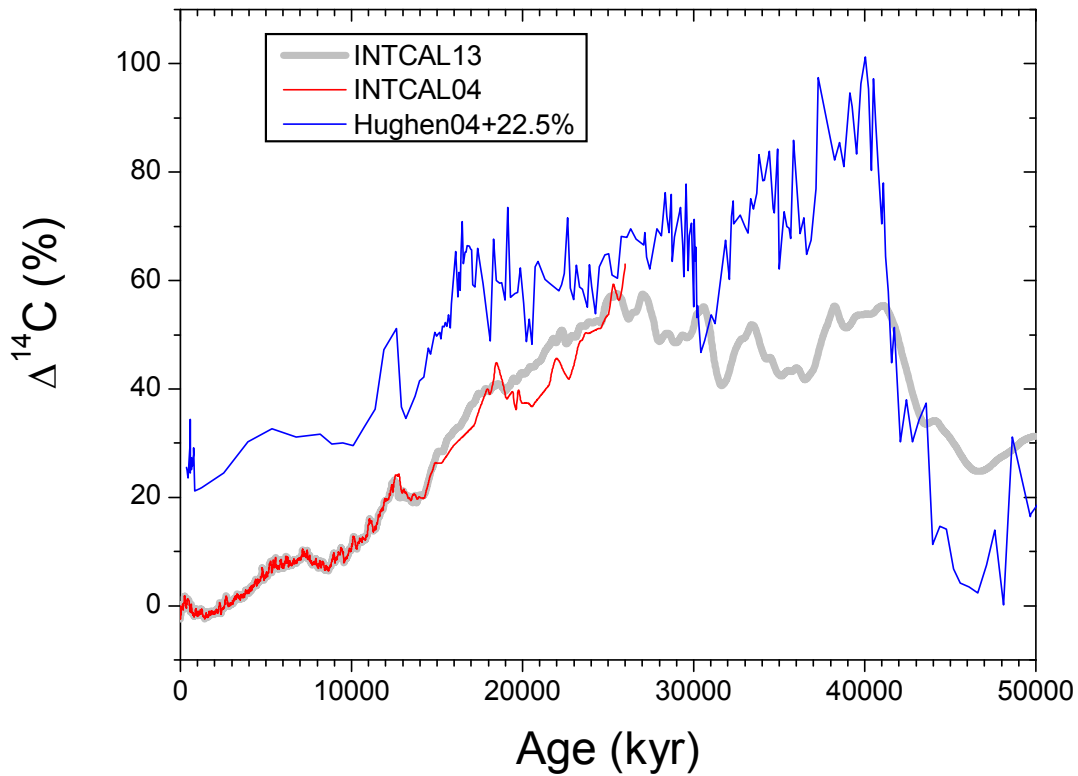

Fig.1. Various series of Δ$^{14}$C: Grey – the most up-to-date series INTCAL13 (Reimer et al., 2013); Red – INTCAL04 series (Reimer et al., 2003); Blue – original series (uplifted by 22.5% following F14) of Cariaco Basin ocean core (Hughen et al., 2004). The series used by F14 is a combination of the red one extended by the blue one past 26 kyrs. In addition, F14 has arbitrarily lifted the entire series by 5% to avoid negative values.

Another reasoning of F14 is that the trend in Δ$^{14}$C during the Holocene is caused by a SN 22 kyr ago. However, the observed Δ$^{14}$C variability during the Holocene is well explained by a combination of solar activity, geomagnetic variability and climate changes (e.g., Solanki et al., 2004; Vomnoos et al., 2006; Usoskin et al., 2007; Snowball & Muscheler, 2007), assuming a constant flux of cosmic rays outside the heliosphere. All the observed variability of Δ$^{14}$C is perfectly explained by these factors without any need to invoke hypothetical supernovae, contrary to F14 claims. This particularly refers to the last 3 kyr (Fig. 6 and Section 2.4 of F14) when the geomagnetic field is very well measured (see Usoskin et al., 2014). If F14 was correct with his reasoning, this would unavoidably lead to reconstructed solar activity that is too low

(even essentially negative) in the early Holocene, which is not observed. On the contrary, the solar activity reconstructed from $^{14}$C shows a tendency to be too high (e.g., Fig. 6 in Vonmoos et al., 2006) suggesting that there was less (contrary to F14's suggestion) $^{14}$C than expected, probably because of changing ocean circulation. Contrary to F14 claims, there is no evidence of historical SNe recorded in cosmogenic isotope data for the last millennium (see Supplement Information, Fig. S2, of Miyake et al., 2013).

Another problem is related to computations of the $^{14}$C production from gamma-rays. F14 uses the yield function (his Fig. 14) of Kovaltsov et al. (2012), but that yield function corresponds to $^{14}$C production by cosmic ray *protons*. F14 explicitly assumes that it can be simply applied to cosmic gammas, but this assumption is wrong, as the physics of the processes induced by high energy protons and gammas in the atmosphere are different. The yield function of atmospheric $^{14}$C production by gammas was calculated by Pavlov et al. (2013, see Fig. 1) and it is much different from the yield function for protons used by F14. Pavlov et al. (2013) said that "the mean yield of $^{14}$C equals to 20-55 atoms erg$^{-1}$ for the gamma-ray flux entering the atmosphere…", while F14 uses about 20000 neutrons erg$^{-1}$ (since production of $^{14}$C is the main sink for neutrons in the atmosphere, this implies roughly the same amount of radiocarbon production by gammas). Lingenfelter & Ramaty (1970) gave the number of ~1000 $^{14}$C atoms erg$^{-1}$ using a very rough estimate, which can serve as an upper limit. Thus, F14 overestimates the $^{14}$C production by orders of magnitude. So,F14 arguments that gamma-ray emission from the SN remnants can produce essential amounts of $^{14}$C are not valid either, as anticipated by Lingenfelter & Ramaty (1970).

The arguments of F14 based on $^{14}$C are invalid because:

(1) They are based on outdated and improperly selected datasets;
(2) They contradict other studies for the Holocene period that explained all the observed variability of $\Delta^{14}$C by solar activity, geomagnetic field and climate;
(3) His computations are based on an improper model, which is not applicable to $^{14}$C production by gammas, leading thus to an error of several orders of magnitude.

*Conclusions*

The high rate and high ionizing photon output claimed by F14 for supernovae in this region of the Galaxy over the last 300 kyr are suspiciously high, and exceed available experimental data. This appears to be because he used obsolete and superseded datasets, and misapplied input parameters for computational models, so that predicted terrestrial $^{14}$C and nitrate deposition exceed correct values by 4 or more orders of magnitude. The case for congruence with data is based on comparison of these incorrect predictions with out-of-date data sets.

We do not dispute indications of a relatively nearby supernova perhaps 2.5 Myr ago (Fields 2004; Bishop 2013; Fry et al. 2014) from $^{60}$Fe deposition. However, the recent work of F14 shows major errors in both interpretation of data and computational modeling.

## *Acknowledgments*

ALM acknowledges the research support of NASA Exobiology. IU and GK acknowledge support from the Academy of Finland (ReSoLVE Centre of Excellence, project No. 272157).